\documentclass[sigconf,review=false]{acmart}

\AtBeginDocument{%
  \providecommand\BibTeX{{%
    \normalfont B\kern-0.5em{\scshape i\kern-0.25em b}\kern-0.8em\TeX}}}

\setcopyright{acmcopyright}
\copyrightyear{2022}
\acmYear{2022}
\acmDOI{10.1145/1122445.1122456}

\acmConference[ICSE '22]{the 44th International Conference on Software Engineering}{May 21--29, 2022}{Pittsburgh, PA}
\acmBooktitle{ICSE '22: the 44th International Conference on Software Engineering,
  May 21--29, 2022, Pittsburgh, PA}
\acmPrice{}
\acmISBN{978-1-4503-XXXX-X/22/05}

\usepackage{listings}
\usepackage{tikz}
\usepackage{todonotes}
\usepackage{xcolor}

\definecolor{mGreen}{rgb}{0,0.6,0}
\definecolor{mGray}{rgb}{0.5,0.5,0.5}
\definecolor{mPurple}{rgb}{0.58,0,0.82}
\definecolor{backgroundColour}{rgb}{0.95,0.95,0.92}

\lstdefinestyle{CStyle}{
    backgroundcolor=\color{backgroundColour},   
    commentstyle=\color{mGreen},
    keywordstyle=\color{magenta},
    numberstyle=\tiny\color{mGray},
    stringstyle=\color{mPurple},
    basicstyle=\ttfamily\footnotesize,
    breakatwhitespace=false,         
    breaklines=true,                 
    captionpos=b,                    
    keepspaces=true,                 
    numbers=left,                    
    numbersep=5pt,                  
    showspaces=false,                
    showstringspaces=false,
    showtabs=false,                  
    tabsize=2,
    language=C
}

\makeatletter
\let\orig@lstnumber=\thelstnumber

\newcommand\lstsetnumber[1]{\gdef\thelstnumber{#1}}
\newcommand\lstresetnumber{\global\let\thelstnumber=\orig@lstnumber}
\makeatother

\newcommand{\examples}{Linux, LLVM, OpenJDK}
\newcommand{\examplelangs}{C, C++, Java}

\begin{document}

\title{XCheck: a Simple, Effective and Extensible Bug Finder using micro-grammar}


\author{Hanwen Zhu}
\affiliation{%
  \institution{McGill University}
  \country{Canada}
}
\email{hanwen.zhu@mail.mcgill.ca}

\author{Junyoung Jang}
\affiliation{%
  \institution{McGill University}
  \country{Canada}
}
\email{junyoung.jang@mail.mcgill.ca}

\author{Xujie Si}
\affiliation{%
  \institution{McGill University}
  \country{Canada}
}
\email{xsi@cs.mcgill.ca}


\begin{abstract}


We propose a simple and effective bug finder, XCheck, 
which is a proof of concept bug finder based on so-called ``micro-grammar''.
The key advantage of XCheck is its extreme simplicity and surprising effectiveness. 
It only consists of a few hundred lines of code but is capable of checking many complicated software systems like \examples{}, which are written in various different languages (e.g., \examplelangs{}). A demo video is available at: https://youtu.be/6FZqs9Un-1U
\end{abstract}




\maketitle

\section{Introduction}

Finding bugs in large complicated software systems has been a grand challenge. Numerous research efforts have been spent in developing capable checkers or testing tools, which can be generally classified into two categories: whitebox approaches and blackbox approaches.
Prominent examples for the whitebox approaches are Saturn~\cite{xie:popl05}, EXE~\cite{cadar:ccs06}, Klee~\cite{cadar:osdi08}, Doop~\cite{bravenboer:oopsla09}, SeaHorn~\cite{gurfinkel:cav15}, SymCC~\cite{poeplau:sec20} which reduce bug finding into symbolic constraint solving, which is then offloaded to various solvers like SAT, SMT, Datalog, or CHC.
Whitebox approaches require carefully modeling the semantics of the source code or their intermediate representations, and the underlying constraint solving process can be extremely expensive.
On the other hand, blackbox approaches like fuzzing~\cite{klees:ccs18,sutton2007fuzzing,serebryany2017oss}, differential testing~\cite{McKeeman98differentialtesting,le:pldi14}, property-based testing~\cite{claessen:icfp00,bulwahn:cpp12,denes:coq14,lampropoulos:oopsla19} avoid modelling semantics and constraint solving but require non-trivial setup to generate and run huge amount of test cases.

In this paper, we consider bug finding from a radically different perspective. All past approaches (whitebox, blackbox, or greybox) can be viewed as discovering concrete and often expensive low-level bug patterns or triggers through either symbolic constraint solving or stochastic search. 
We advocate finding bugs in an abstract and lightweight high-level perspective. Approaches in this perspective are ``cheap-and-effective by design''.
For instance, the entire checker only consists of a few hundred lines of code but is sufficient to check many complicated software systems like \examples{}, which are written in various different languages (e.g., \examplelangs{}).
There are very few relevant research work in this perspective, and we are mainly inspired by the concept of ``micro-grammar'' proposed by \citet{brown:asplos16}.
The key insight is to abstract away irrelevant programming details through abstracting away language details. 
Many high-level bug patterns can be effectively captured by a tiny fraction of the language that is used to implement the software. 
This insight enables extending a checker to check a new software written in a new programming language in just a few lines of code. 

We are excited about evaluating this insight as well as promoting this new perspective in general. However, there is no open-source realization of the micro-grammar idea, which to our best knowledge is only implemented in a commercial product. Our key motivation is to develop an open-source implementation, which we call XCheck, and evaluates its effectiveness. 

Interestingly, our evaluation finds a bug of the bugs reported by \citet{brown:asplos16}. This is possible because XCheck disagrees with one claim made in the original paper, and our further investigation indicates XCheck is right while the original paper is wrong. We hope XCheck could serve as a valuable basis for other practical bug finding tools, as well as many future works in this line of research. 

The rest of the paper is organized as follows. Section 2 gives a brief overview of the implementation of XCheck, and section 3 highlights several examples used in our evaluation. 
Section 4 discusses related work, while section 5 concludes and shares some extensions of XCheck in the near future.

\section{XCheck Framework}
We use \lstinline{javaStyle} from the Haskell Parsec library to build the lexers and then build parser combinators on top of the lexers. Parser combinators are easily composable, so we can implement a new parser by combining parser combinators.

\subsection{Parser}
Our parser has the following distinctive features: a micro-grammar based abstract syntax tree(AST) and two-step parsing.

\subsubsection{Micro-Grammar based AST}
The micro-grammar only captures an incomplete representation of a language~\cite{brown:asplos16}. For example, consider the following code snippet:
\begin{lstlisting}
if (x >= 3) foo();
while (x <= 2) x++;
\end{lstlisting}

Two parse trees would be generated according to the micro-grammar, one for the \textit{if} statement, and the other for the \textit{while} statement. The micro-grammar does not have sophisticated production rules for the expressions. All expressions are left uninterpreted as \textit{wildcards}, which we discuss in detail later. The micro-grammar would describe the above example as:
\begin{lstlisting}
if ["x",">=","3"] ["foo","(", ")"]
while ["x","<=","2"] ["x","++"]
\end{lstlisting}

In our implementation, we divide a language into two parts: \textit{statements} and \textit{expressions}. In the above example, \lstinline{while} is a statement and \lstinline{x <= 2} is an expression. We make production rules for all the control-flow statements, for example, \lstinline{for}, \lstinline{while}, etc. The irrelevant parts of a language that we believe to be bug-free, for example, \textit{struct} definition in C language, would be treated as a terminal, and we skip them when finding the bugs. We include all the expressions in a language. There is currently an extra \textit{wildcard} expression. A wildcard is simply a list of valid tokens in a language. To make our parser as portable as possible, ideally the entire source code should be treated as a wildcard, we then apply the \textit{sliding-window} technique to parse the wildcard token by token. If the current window is a valid statement in a language, the parser will replace the wildcard with a parse tree, otherwise the parser moves forward by one token.

To facilitate building the checkers, we make statement and expression instances of \lstinline{Eq} and \lstinline{Ord}. Two statements are equal if they have the same constructor, and the expressions they contain are all equal; two expressions are equal if they have the same constructor, and the atomic values they contain are all equal.

\subsubsection{Two-step Parsing}

Our parser parses a file in two steps. In the initial step, the parser parses a file according to the micro-grammar so that it would produce a result with only statements plus wildcard expressions. There are three important parsers that are used in this step: \lstinline{skipTo}, \lstinline{anyTkn} and \lstinline{balancedP}.
\lstinline{anyTkn} parses any valid token of a language. Some valid inputs are ambiguous. For example, given an input string "++", \lstinline{anyTkn} might wrongly parse it to "+". To avoid ambiguity, we sort the valid tokens by the length in descending order, so \lstinline{anyTkn} will try the longer valid tokens before the short ones. \lstinline{skipTo} takes as input a suffix \lstinline{P}, usually the end of statement symbol of a language, and uses \lstinline{anyTkn} to parse a string to a wildcard expression up to \lstinline{P}. \lstinline{balancedP} uses \lstinline{anyTkn} and parses a parentheses-enclosed string by ensuring that there is an equal amount of left parentheses and right parentheses. If we naively use \lstinline{skipTo(")")}, we might get a wrong result because there might be other right parentheses within the outermost parentheses. 

After the first step, we use an expression  parser to parse those wildcard expressions. The expression parser should never fail, since those expressions are where bugs are possibly located. The parsed result is represented by a list of parse trees for the statements. We then pattern match with statements and expressions to analyze the code and find bugs.

\subsection{Checker}
We implement four checkers to find the potential logical errors and null pointer dereferencing:
\begin{itemize}
    \item \textbf{redundant logical conditions and redundant branches} detects \lstinline{if}, \lstinline{case} statements with repeated condition/branches. These two checkers work in a similar manner. They pattern match against \lstinline{if} and \lstinline{switch} statements and compare all the subsequent \lstinline{else if} or \lstinline{case} statements to determine if there are redundant ones.
    \item \textbf{suspicious loop ending conditions} detects \lstinline{for} statements with suspicious ending conditions. This checker examines if the comparison symbol in the ending condition of a \lstinline{for} loop contradicts the update expression. For instance, this checker produces a warning when the ending condition contains "<" and the update expression contains "--".
    \item\textbf{null pointer dereferencing} detects possible null pointers dereferencing by tracing the state of pointers. We divide pointers into two groups, the \textit{nonnull} pointers and the \textit{nullable} pointers. If a pointer is dereferenced, it implies that it is nonnull; if a pointer is used alone as a logical expression,  it implies that this pointer is nullable. In this checker, we use a list denoted by $L$ to record the non-null pointers in the code. If an element in $L$ appears in an assignment expression later, we remove it from $L$. When the checker encounters a null pointer check "\lstinline{if}($p$)" and $p \in L$, the checker would produce a warning.
\end{itemize}

We use \lstinline{setPosition} and \lstinline{getPosition} utilities from the Parsec library in the checkers to identify the exact positions of bugs in the code.

\section{Evaluation}

We tested our checker against Linux5.15, OpenJDK8 and LLVM3.5.1 and successfully found issues in these projects. We also identified a mistake in the paper proposed by \citet{brown:asplos16}.

\subsection*{null pointer dereferencing in OpenJDK and LLVM}

We found instances of null pointer dereferencing in CipherCore.java from OpenJDK and InstCombineAddSub.cpp from LLVM:
\begin{lstlisting}[style=Cstyle, firstnumber=885]
/* jdk/src/share/classes/com/sun/crypto/provider/CipherCore.java */
int outputCapacity = output.length - outputOffset;
  ...
if ((output == null) || (outputCapacity < minOutSize)) {
      ...
}
\end{lstlisting}
\begin{lstlisting}[style=Cstyle, firstnumber=455, mathescape=true]
/* llvm/lib/Transforms/InstCombine/InstCombineAddSub.cpp */
Value *Opnd0_0 = I0->getOperand(0);$\lstsetnumber{\ldots}$
  ...$\lstresetnumber\setcounter{lstnumber}{489}$
if (I0) Flags &= I->getFastMathFlags();
\end{lstlisting}

We checked the entire CipherCore.java file and line 440 to line 516 of InstCombineAddSub.cpp. In InstCombineAddSub.cpp, pointer \lstinline{I0} is dereferenced in line 456; in CipherCore.java, object \lstinline{output} is dereferenced in line 886. Notice the difference here: Java uses "." to indicate pointer dereferencing, but C++ uses "->". To migrate the C++ checker to Java, we simply change how the checker finds dereferenced pointers from tracing "->" to tracing ".", and the rest of the checker remains the same. Both checkers identified where the pointers are checked against \lstinline{null} (line 888 for the Java example; line 490 for the C++ example) and emit an error.

\subsection*{false positive result when checking Linux from previous work}

We checked object.c from Linux using our C checker and did not find any issue. This result contradicts with the result mentioned in the previous paper~\cite{brown:asplos16}. In the previous paper, the authors mentioned that there is null pointer dereferencing in line 233 in object.c because function pointer \lstinline{state->work} is dereferenced in line 233 and checked against \lstinline{null} in line 250:
\begin{lstlisting}[style=Cstyle, firstnumber=232, mathescape=true]
/* linux/fs/fscache/object.c */
new_state = state->work(object, event);$\lstsetnumber{\ldots}$
  ...$\lstresetnumber\setcounter{lstnumber}{247}$
object->state = state = new_state;

if (state->work) {$\lstsetnumber{\ldots}$
      ...$\lstresetnumber\setcounter{lstnumber}{255}$
}
\end{lstlisting}

\lstinline{state->work} is indeed used in a \lstinline{if} statement in line 250, but actually a new value \lstinline{new_state} is assigned to \lstinline{state} in line 248. Thus, the checker should remove \lstinline{state->work} from the nonnull list and not emit an error. In our C checker, we implement a helper method to find the \textit{root} of a pointer dereferencing expression. When a pointer \lstinline{p} is found in an assignment expression, this helper method removes all elements in the non-null list that are initially accessed from \lstinline{p}, for example, \lstinline{p->q} and \lstinline{p->l->k}. Therefore, our C checker do not report null pointer dereferencing for \lstinline{state}.

\section{Related works}
Our implementation is based on micro-grammars introduced by \citet{brown:asplos16}. Their approach simplifies static program analysis and allows its application to scale to larger systems such as Linux and Firefox. However, as discussed in the last section, their null pointer dereferencing checker has a bug in the handling of variable assignments.

Other static program analyzers based on concrete language specifications such as PREfast~\cite{ball:ifm04}, ESP~\cite{das:pldi02}, ESC~\cite{flanagan:pldi02} depend on specific implementations of their target languages. This dependence allows those analyzers to exploit some implementation-specific features so that they can detect more target-specific bugs. However, because of the complexities of their targets, they are often limited to their original target and hard to be ported into other implementations, let alone other languages.

For implementation/language agnostic static analysis framework, \citet{xie:popl05} presented Saturn framework operating on a small language, scalar. However, their approach requires a full language parser and transpiler from the language to scalar, which increases the complexity of adaptation.

Dynamic program analyzers such as Taint~\cite{newsome:ndss05}, EXE~\cite{cadar:ccs06}, Valgrind~\cite{nethercote:pldi07}, and Dytan~\cite{clause:issta07} does not require a full language specification for their analysis. However, as their analysis depends on machine instructions, cross-machine port of such analyzers requires considerable works.

SymCC~\cite{poeplau:sec20} and klee~\cite{cadar:osdi08} analyze LLVM IR, and thus, they resolved the issue of language and machine agnostic. However, they use symbolic executions for their analysis, which suffer from their path explosion problems.

\section{Conclusion and Future work}
With micro-grammars, we implemented a powerful yet simple and portable checker, XCheck, that can find some possible bugs in real world examples such as \examples{}. This simplicity and portability are exactly due to the micro-grammar approach. It allows the parser combinators of XCheck to remain simple and to be reused across languages.

We plan to improve XCheck by porting it to other languages that are much different from C/C++, for example Python, and adding more checkers like checking suspicious \lstinline{while} loop condition. Besides that, we will boost the robustness of our parser by implementing the sliding-window technique.

\bibliography{reference}

\end{document}